# Fault Tolerant Variable Block Carry Skip Logic (VBCSL) using Parity Preserving Reversible Gates

Md. Saiful Islam, M. M. Rahman*, Zerina Begum, and M. Z. Hafiz

Institute of Information Technology, University of Dhaka, Dhaka-1000, Bangladesh
*Dept. of Computer Science and Engineering, American International University-Bangladesh, Dhaka-1213, Bangladesh
Email: saiful@iit.du.ac.bd, mus_mahbub@hotmail.com, zerin@univdhaka.edu, jewel@univdhaka.edu

*Abstract*—Reversible logic design has become one of the promising research directions in low power dissipating circuit design in the past few years and has found its application in low power CMOS design, digital signal processing and nanotechnology. This paper presents the efficient design approaches of fault tolerant carry skip adders (FTCSAs) and compares those designs with the existing ones. Variable block carry skip logic (VBCSL) using the fault tolerant full adders (FTFAs) has also been developed. The designs are minimized in terms of hardware complexity, gate count, constant inputs and garbage outputs. Besides of it, technology independent evaluation of the proposed designs clearly demonstrates its superiority with the existing counterparts.

## I. INTRODUCTION

Irreversible logic circuits dissipate heat in the amount of *kT* ln2 Joule for every bit of information that is lost irrespective of their implementation technologies, where *k* is the Boltzmann constant and *T* is the operating temperature [1]. Information is lost when the circuit implements nonbijective functions. Therefore, in irreversible logic circuit the input vector cannot be recovered from its output vectors. Reversible logic circuit by definition realizes only those functions having one-to-one mapping between its input and output assignments. Hence in reversible circuits no information is lost. According to [2] zero energy dissipation would be possible only if the network consists of reversible gates. Thus reversibility will become an essential property in future circuit design.

Reversible logic imposes many design constraints that need to be either ensured or optimized for implementing any particular Boolean functions. Firstly, in reversible logic circuit the number of inputs must be equal to the number of outputs. Secondly, for each input pattern there must be a unique output pattern. Thirdly, each output will be used only once, that is, no fan out is allowed. Finally, the resulting circuit must be acyclic [3-5]. Any reversible logic design should minimize the followings [6]:

- **Garbage Outputs (GO)**: outputs that are not used as primary outputs are termed as garbage outputs
- **Constant Inputs (CI)**: constants are the input lines that are either set to zero(0) or one (1) in the circuit's input side
- **Gate Count (GC)**: number of gates used to realize the system
- **Hardware Complexity (HC)**: refers to the number of basic gates (NOT, AND and EXOR gate) used to synthesize the given function

In addition of the above four, the following two measures can be taken into consideration in designing a digital system:
- **Area Consumption (AC)**: the area consumed by a design can be calculated as the sum of the width of individual gates used to realize the system.

$$AC = \sum_{i=1}^{n} w_i$$

- **Path Delay (PD)**: total number of gates on the way of final output.

Parity checking is one of the widely used mechanisms for detecting single level fault in communication and many other systems. It is believed that if the parity of the input data is maintained throughout the computation, no intermediate checking would be required [7-8]. Therefore, parity preserving reversible circuits will be the future design trends towards the development of fault tolerant reversible systems in nanotechnology. A gating network will be parity preserving if its individual gates are parity preserving [7]. Thus, we need parity preserving reversible logic gates to construct parity preserving reversible circuits.

This paper presents efficient approaches for designing fault tolerant carry skip adder using modified IG gate (MIG) and its basic building block FTFA proposed in [9-10] [17]. The proposed carry skip adder is optimized in terms of gate count, garbage outputs and constant inputs and offers less hardware complexity. Variable block carry skip logic using FTFA circuits has also been given. Finally, technology independent evaluation of the proposed carry skip adders has been presented.

## II. REVERSIBLE LOGIC GATES

A gate that implements any bijective function involving $n$ inputs and $n$ outputs is called an $n*n$ reversible logic gate. There exist many reversible gates in the literature. Among them 2*2 Feynman gate [11] (shown in Fig. 1), 3*3 Fredkin gate [12] (shown in Fig. 2), 3*3 Toffoli gate [13] (shown in Fig. 3) and 3*3 Peres gate [14] (shown in Fig. 4) are the most referred. Feynman (FG), Fredkin (FRG) and Peres (PG) gates are one through gates, that is, one of its output lines is identical to one of its input lines. On the other hand, Toffoli gate is two through, that is, two of its outputs are identical to two of its inputs. It can be easily verified that all of these gates are reversible. Each gate has an equal number of input and output lines. For each input combination there is a unique output combination.

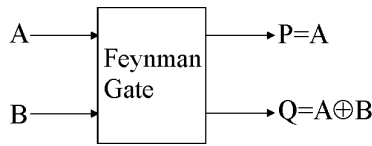

Fig. 1(a) 2*2 Feynman gate

| A | B | P | Q |
|---|---|---|---|
| 0 | 0 | 0 | 0 |
| 0 | 1 | 0 | 1 |
| 1 | 0 | 1 | 1 |
| 1 | 1 | 1 | 0 |

Fig. 1(b) Truth table of Feynman gate

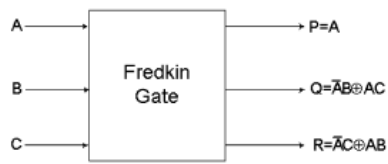

Fig. 2(a) 3*3 Fredkin gate

| A | B | C | P | Q | R |
|---|---|---|---|---|---|
| 0 | 0 | 0 | 0 | 0 | 0 |
| 0 | 0 | 1 | 0 | 0 | 1 |
| 0 | 1 | 0 | 0 | 1 | 0 |
| 0 | 1 | 1 | 0 | 1 | 1 |
| 1 | 0 | 0 | 1 | 0 | 0 |
| 1 | 0 | 1 | 1 | 1 | 0 |
| 1 | 1 | 0 | 1 | 0 | 1 |
| 1 | 1 | 1 | 1 | 1 | 1 |

Fig. 2(b) Truth table of Fredkin gate

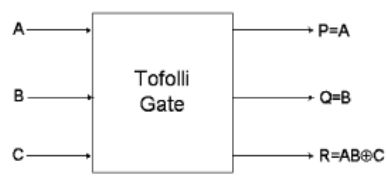

Fig. 3(a) 3*3 Toffoli gate

| A | B | C | P | Q | R |
|---|---|---|---|---|---|
| 0 | 0 | 0 | 0 | 0 | 0 |
| 0 | 0 | 1 | 0 | 0 | 1 |
| 0 | 1 | 0 | 0 | 1 | 0 |
| 0 | 1 | 1 | 0 | 1 | 1 |
| 1 | 0 | 0 | 1 | 0 | 0 |
| 1 | 0 | 1 | 1 | 0 | 1 |
| 1 | 1 | 0 | 1 | 1 | 1 |
| 1 | 1 | 1 | 1 | 1 | 0 |

Fig. 3(b) Truth table of Toffoli gate

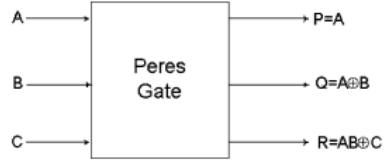

Fig. 4(a) 3*3 Peres gate

| A | B | C | P | Q | R |
|---|---|---|---|---|---|
| 0 | 0 | 0 | 0 | 0 | 0 |
| 0 | 0 | 1 | 0 | 0 | 1 |
| 0 | 1 | 0 | 0 | 1 | 0 |
| 0 | 1 | 1 | 0 | 1 | 1 |
| 1 | 0 | 0 | 1 | 1 | 0 |
| 1 | 0 | 1 | 1 | 1 | 1 |
| 1 | 1 | 0 | 1 | 0 | 1 |
| 1 | 1 | 1 | 1 | 0 | 0 |

Fig. 4(b) Truth table of Peres gate

### III. PARITY PRESERVING REVERSIBLE GATES

A reversible gate will be parity preserving if the parity of the inputs matches the parity of the outputs. Mathematically, a reversible gate having $n$ input lines and $n$ output lines will be parity preserving if and only if:

$$I_1 \oplus I_2 \oplus \cdots \oplus I_n \leftrightarrow O_1 \oplus O_2 \oplus \cdots \oplus O_n$$

where $I_i$ and $O_j$ are the input and output lines. Not all of the gates presented in section II are parity preserving.

Only Fredkin gate (FRG) is parity preserving and it can be easily verified by examining its truth table. That is, Fredkin gate maintains $A \oplus B \oplus C \leftrightarrow P \oplus Q \oplus R$. It can also be said that the Fredkin gate is zero preserving (once preserving as well) and therefore conservative [15]. Other parity preserving reversible logic gates are 3*3 Feynman Double gate [7] (shown in Fig. 5), 3*3 New Fault Tolerant gate [16] (shown in Fig. 6) and newly proposed 4*4 IG gate [9-10] (shown in Fig. 7). Feynman Double gate can be as used as the fault tolerant copying gate when it's B and C input lines are set to constants. The first three output lines of IG gate produce the same output as PG gate. The fourth one can be considered as garbage if we wish to replace PG by IG. The fourth output line of IG gate can also be minimized to reduce the hardware complexity in [17] as follows:

$$BD \oplus B'(A \oplus D) \Leftrightarrow AB' \oplus D$$

The modified version of IG (that is, MIG gate) is depicted in Fig. 8.

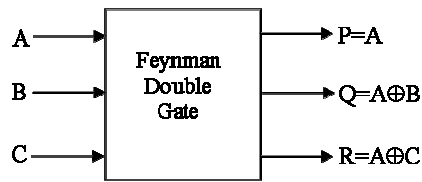

Fig. 5 3*3 Feynman Double gate

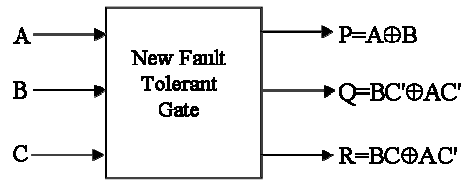
Fig. 6 3*3 New Fault Tolerant gate

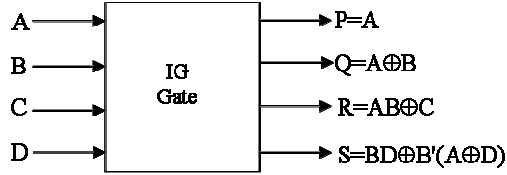
Fig. 7 4*4 IG gate

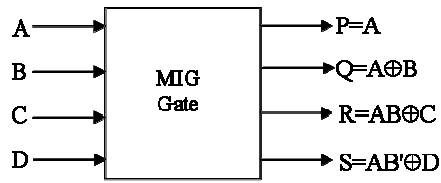
Fig. 8 4*4 MIG gate

## IV. SYNTHESIS OF FAULT TOLERANT REVERSIBLE CARRY SKIP ADDER

The basic building block of many complex computational systems is the full adder (FA). Realization of the efficient reversible full adder circuit given in [3-5] includes two 3*3 Peres gates (shown in Fig. 9). The circuit is minimized in terms of gate count, garbage outputs, constant inputs and hardware complexity. It has also been proved in [3-5] that a reversible full adder circuit can be realized with at least two garbage outputs and one constant input.

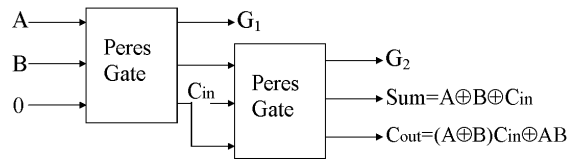
Fig. 9 Reversible full adder circuit given in [3-5]

Fault tolerant logic synthesis of reversible full adder circuit requires that its individual gate unit must be fault tolerant reversible gates. It has been proved in [9-10] that a fault tolerant reversible full adder circuit requires at least three garbage outputs and two constant inputs. This paper will use MIG gate for fault tolerant reversible full adder implementation as used in [17] and will thereby minimize the hardware complexity of the presented systems. The Fig. 10 shows a 1-bit fault tolerant reversible full adder realized using two MIG gates.

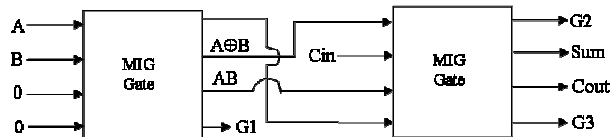
Fig. 10 Fault tolerant reversible full adder circuit includes two 4*4 MIG gates

The block diagram of the fault tolerant reversible full adder can be depicted as follows:

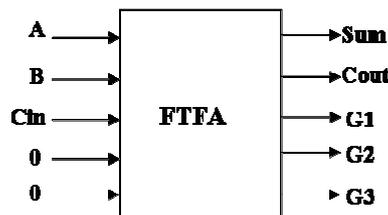
Fig. 11 Block diagram of fault tolerant full adder

The most straightforward realization of a final stage adder for two N-bit operands is ripple carry adder (RCA). The RCA requires N full adders (FAs). The carry out of the $i^{th}$ FA is connected to the carry in of the $(i+1)^{th}$ FA. The Fig. 12 shows a reversible logic implementation of an N-bit final stage fault tolerant ripple carry adder.

A carry-skip adder reduces the carry-propagation time by skipping over groups of consecutive adder stages. The carry-skip adder is usually comparable in speed to the carry look-ahead technique, but it requires less chip area and consumes less power [15]. The Fig. 13 shows a reversible logic implementation of a 4-bit fault tolerant carry skip adder block proposed in [17]. The block carry input $c_{in}$ is propagated as the block carry output $c_{out}$, if the block propagate P is one. The carry skip adder worst case delay is the when the carry generated in the very first full adder ripples through all full adder stages in the first block, generates $c_{out}$ for the first block, skips all the intermediate blocks, and ripples through the full adder stages of the last block.

This paper presents two new fault tolerant reversible carry skip adders, shown in Fig. 14 and Fig. 15. The carry skip logic in Fig. 14 has been developed using 3 NFTs and 1 FRG instead of 4 NFTs and 1 F2G as in [17]. The carry skip logic in Fig. 15 uses 4 FRGs as in [15] and 1 F2G to avoid the fanout.

A *B* bit FTFA requires 2B MIG using the circuit in Fig. 10. The *B* input AND gate requires *B*-1 NFTs in Fig. 14. Therefore, a *B* bit carry skip adder block requires 2B MIG, B-1 NFTs, 1 FRG and 1 F2G using the FTFA circuit in Fig. 10.

$$d_{c_{ripple}}(B) = B + 3 \quad (1)$$

Consider the *B*-bit carry skip adder in Fig. 14 generating a block carry out $c_{out}$ via carry ripple through the FTFAs. The least significant FTFA will require a path delay of 2 MIGs to generate $c_1$ from the addends. Then the carry "ripples" through the subsequent FTFAs with a path delay of 1 MIG per bit. Finally, $c_{out}$ is generated by the FRG in the left of Fig. 14. Therefore, the delay to generate block carry out $c_{out}$ (via ripple) with a B bit carry skip adder block is:

$$d_{c_{skip}}(B) = \lceil \log_2^B \rceil + 4 \quad (2)$$

The total (worst case) delay $T_{fixed}$ of an N bit carry skip adder with fixed block size B is the sum of the ripple carry delay through the first carry skip adder block, the carry skip delays through the intermediate blocks and the ripple carry delay through the last block, or

$$T_{fixed} = (B+3) + \left(\frac{N}{B} - 2\right)\left(\lceil \log_2^B \rceil + 4\right) + (B+3) \quad (3)$$

However, the assumption $\lceil \log_2^B \rceil \approx \frac{B}{2}$ is valid for the small block sizes B applicable to carry skip adder designs. Thus, (3) can be rewritten as

$$T_{fixed} = B + \frac{N}{2} + \frac{4N}{B} - 2 \quad (4)$$

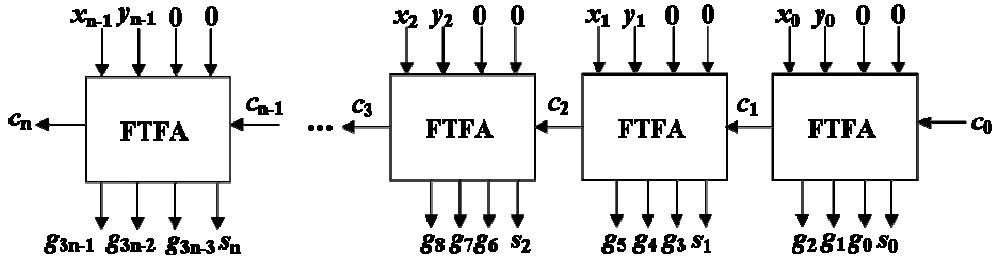

Fig. 12 Reversible logic implementation of fault tolerant *n*-bit ripple carry adder [17]

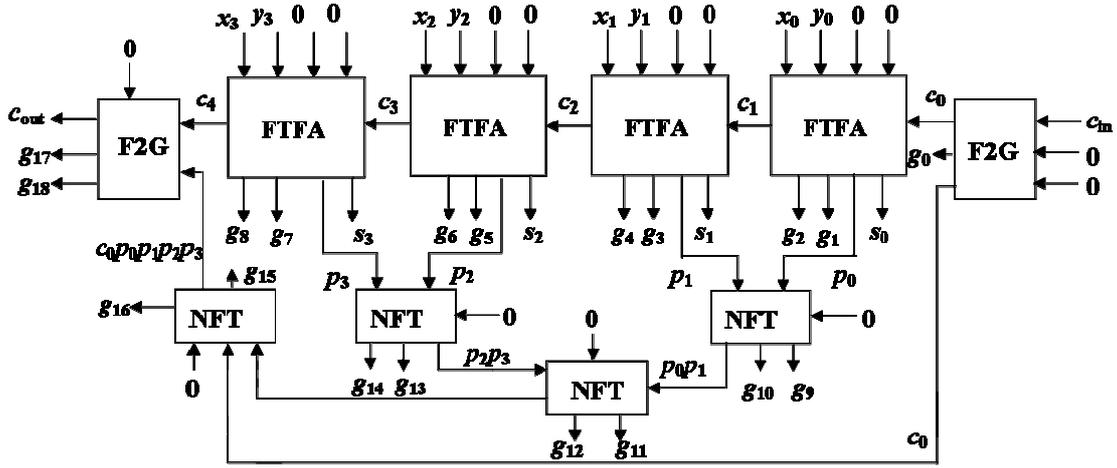

Fig. 13 Reversible logic implementation of fault tolerant 4-bit carry skip adder [17]

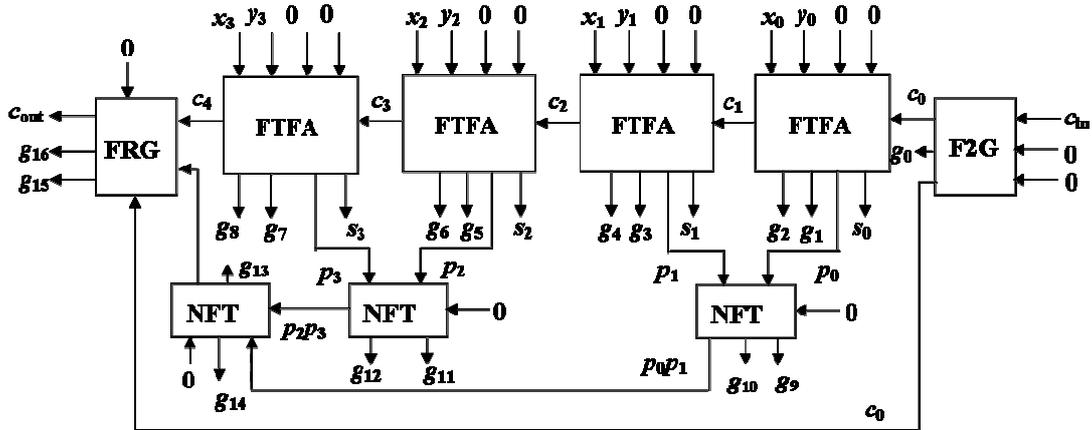

Fig. 14 Proposed fault tolerant reversible 4-bit carry skip adder, CSL using NFTs

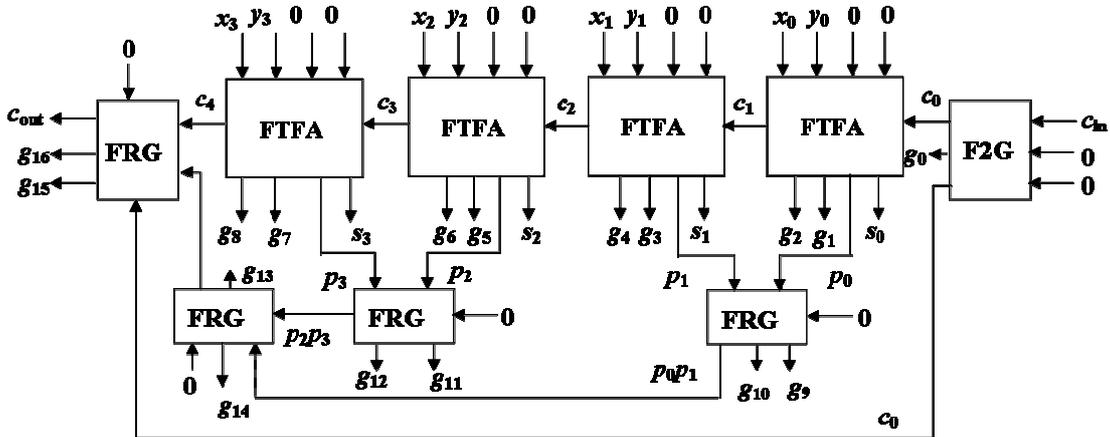

Fig. 15 Proposed fault tolerant reversible 4-bit carry skip adder, CSL using FRGs

Minimizing $T_{fixed}$ with respect to block size B gives

$$1 - \frac{4N}{B^2} = 0 \quad \text{or} \quad B_{opt} = \sqrt{4N} \quad (5)$$

Substituting (5) into (4) gives

$$T_{fixed} = \frac{N}{2} + 4\sqrt{N} - 2 \quad (6)$$

The calculation of $T_{fixed}$ for the carry skip adder shown in Fig. 15 is same. The only difference between these two designs is in the CSL implementations. The CSA in Fig. 15 uses FRGs instead of NFTs to implement B-input AND gate.

## I. VARIABLE BLOCK CARRY SKIP ADDER

Varying the size of the carry skip blocks can reduce the total worst case delay, since carries generated or absorbed in the adder center have shorter data paths [15]. Without loss of generality, the first and last carry skip blocks are b bits wide, and the carry skip adder is divided into *t* blocks, where t is even. Let's assume that carry skip block sizes are:

$$b \quad b+1 \quad \cdots \quad b+\frac{t}{2}-1 \quad b+\frac{t}{2}-1 \quad \cdots \quad b+1 \quad b \quad (7)$$

Summing the number of bits in the blocks, equating to N, and rearranging gives

$$b = \frac{N}{t} - \frac{t}{4} + \frac{1}{2} \quad (8)$$

The total (worst case) delay $T_{variable}$ of an N bit carry skip adder with the variable block size is the sums of the ripple carry delay through the first carry skip adder block, the carry skip delays through the intermediate blocks and the ripple carry delay through the last block. Assuming the variable block sizes in (7), the total delay is:

$$T_{variable} = 2(b+3) + 2\sum_{k=b+1}^{b+\frac{t}{2}-1}\left(\lceil \log_2^k \rceil + 4\right) \quad (9)$$

Again assuming $\lceil \log_2^k \rceil \approx \frac{k}{2}$ and rearranging gives

$$T_{variable} = 2b - 2 + 4t + \sum_{k=b+1}^{b+\frac{t}{2}-1} k \quad (10)$$

Using the identity

$$\sum_{k=X}^{Y} K = \frac{Y(Y+1) - X(X-1)}{2}$$

To simplify (10) yields

$$T_{variable} = 2b - 2 + 4t + \frac{\left(b+\frac{t}{2}-1\right)\left(b+\frac{t}{2}\right) - (b)(b+1)}{2} \quad (11)$$

$$= \frac{t^2}{8} + \frac{15t}{4} + \frac{bt}{2} + b - 2$$

Inserting (8) into (11) and collecting terms gives

$$T_{variable} = \frac{15t}{4} - \frac{3}{2} + \frac{N}{t} + \frac{N}{2} \quad (12)$$

The optimal number of blocks is found with

$$\frac{\partial T}{\partial t} = \frac{15}{4} - \frac{N}{t^2} = 0 \quad \text{or} \quad t_{opt} = \frac{2}{\sqrt{15}}\sqrt{N} \quad (13)$$

Therefore, the optimal block size carry skip adder has delay

$$T_{variable} = \frac{N}{2} + \sqrt{15}\sqrt{N} - \frac{3}{2} \quad (14)$$

## II. RESULT AND DISCUSSION

The presented carry skip adder circuits can be evaluated in terms of hardware complexity, gate count, constant inputs and garbage outputs produced. Let

$\alpha$ = A two input EXOR gate calculation
$\beta$ = A two input AND gate calculation
$\delta$ = A NOT gate calculation
T = Total logical calculation

The total logical calculation of the CSA presented in [15] and [17] are T= $48\alpha+96\beta+24\delta$ and T=$40\alpha+28\beta+12\delta$ respectively. On the contrary, the total logical calculation of the presented CSAs in this study are T=$37\alpha+29\beta+12\delta$ (Fig. 14) and T=$34\alpha+32\beta+5\delta$ (Fig 15). The presented CSAs require only 13 reversible gates whereas the CSA designs in [15] and [17] require 14 and 24 reversible gates respectively. The presented designs also significantly reduce the required number of constant inputs and garbage outputs. It should also be noted that the area consumptions of the presented designs are minimum compared to the designs found in the literature [15] [17]. Evaluation of the proposed designs can be comprehended easily with the help of the comparative results given in Table I.

The optimum number of block size, $B_{opt}$, of the presented CSAs is 8 for *N*=16. On the contrary, the number of blocks for the designs in [15] and [17] are fractional and therefore impractical (Block size must be integer). Fig. 16 shows the adder size and the corresponding delay for different CSA implementations including those presented in this study. For N ≥ 16, the proposed CSAs show better performance than the existing counterparts.

## III. CONCLUSION

This paper presents the efficient approaches for designing variable block skip logic using parity preserving reversible gates that are minimized in terms of hardware complexity, gate count, constant inputs and garbage outputs. Technology independent evaluation demonstrates its superiority with the existing designs. The two new measures area consumption (AC) and path delay (PD) has been taken into consideration to evaluate all designs including those presented in this paper.

## REFERENCES

[1] R. Landauer, "Irreversibility and heat generation in the computing process", IBM J. Research and Development, vol. 5, pp. 183-191, 1961.
[2] C. H. Bennet, "Logical reversibility of computation", IBM J. Research and Development, vol. 17, no. 6, pp. 525-532, 1973.
[3] M. S. Islam, M. R. Islam, M. R. Karim, A. A. Mahmud and H. M. H. Babu, "Variable block carry skip logic using reversible gates", In Proc. of 10th International Symposium on Integrated Circuits, Devices & Systems, Singapore, pp 9-12, 8-10 September, 2004.
[4] M. S. Islam, M. R. Islam, M. R. Karim, A. A. M. and H. M. H. Babu, "Minimization of adder circuits and variable block carry skip logic using reversible gates", In Proc. of 7th International Conference on Computer and Information Technology, Dhaka, Bangladesh, 26-28 December, 2004, pp. 378-383.
[5] M. S. Islam and M. R. Islam, " Minimization of reversible adder circuits" , Asian Journal of Information Technology, vol. 4, no. 12, pp. 1146-1151, 2005.

TABLE I COMPARATIVE EXPERIMENTAL RESULT OF DIFFERENT FAULT TOLERANT CSA CIRCUITS

| CSA Design | Gate Count | Hardware Complexity | Constant Inputs | Garbage Outputs | Area Consumption |
|---|---|---|---|---|---|
| 1-bit FTFA [17] | 2 MIG | $6\alpha+4\beta+2\delta$ | 2 | 3 | 8 |
| 1-bit FTFA [9][10] | 2 IG | $8\alpha+6\beta+2\delta$ | 2 | 3 | 8 |
| 1-bit FTFA [15] | 4 FRG | $8\alpha+16\beta+4\delta$ | 2 | 3 | 15 |
| 4-bit CSA: This study (Fig. 14) | 8MIG+3NFT+1F2G+1FRG=13 | $37\alpha+29\beta+12\delta$ | 14 | 17 | 47 |
| 4-bit CSA: This study (Fig. 15) | 8 MIG+ 4 FRG+ 1F2G =13 | $34\alpha+32\beta+5\delta$ | 14 | 17 | 47 |
| 4-bit CSA[17] | 8 MIG + 4NFT +2 F2G=14 | $40\alpha+28\beta+12\delta$ | 15 | 19 | 50 |
| 4-bit CSA [15] | 24 FRG | $48\alpha+96\beta+24\delta$ | 19 | 24 | 72 |

TABLE II   $T_{fixed}$ AND $T_{variable}$ OF DIFFERENT CSA IMPLEMENTATIONS

| CSA Presented | $T_{fixed}$ | $B_{opt}$ | $T_{variable}$ | $t_{opt}$ |
|---|---|---|---|---|
| This Study | $\frac{N}{2}+4\sqrt{N}-2$ | $\sqrt{4N}$ | $\frac{N}{2}+\sqrt{15}\sqrt{N}-\frac{3}{2}$ | $\frac{2}{\sqrt{15}}\sqrt{N}$ |
| CSA [17] | $\frac{N}{2}+4.47\sqrt{N}-4$ | $\sqrt{5N}$ | $\frac{N}{2}+\sqrt{19}\sqrt{N}-\frac{7}{2}$ | $\frac{2}{\sqrt{19}}\sqrt{N}$ |
| CSA [15] | $\frac{N}{2}+7.75\sqrt{N}-4$ | $\sqrt{\frac{5N}{3}}$ | $\frac{N}{2}+\sqrt{51}\sqrt{N}-\frac{5}{2}$ | $\frac{2\sqrt{51}}{17}\sqrt{N}$ |

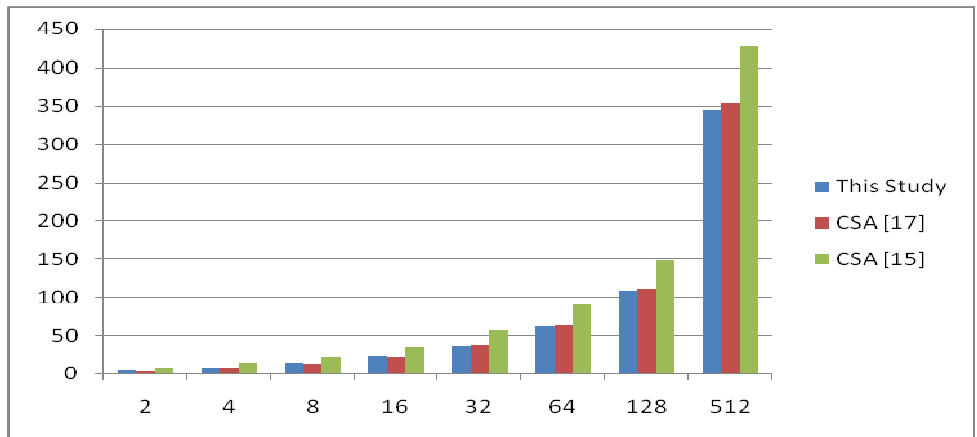

Fig. 16 Adder Size vs. Delay for different CSA implementations: Adder Size is placed on horizontal axis and delay is placed on vertical axis

[6]  M. S. Islam, M. M. Rahman, Z. Begum and M. Z. Hafiz, "Low cost quantum realization of reversible multiplier circuit", Information Technology Journal, vol. 8, no. 2, pp.208-213, 2009.
[7]  B. Parhami , "Fault tolerant reversible circuits", in Proceedings of 40th Asimolar Conf. Signals, Systems, and Computers, Pacific Grove, CA, pp. 1726-1729, October 2006.
[8]  R. K. James, Shahana T. K., K. P. Jacob and S. Sasi, "Fault Tolerant Error Coding and Detection using Reversible Gates", IEEE TENCON, pp. 1- 4, 2007.
[9]  M. S. Islam, M. M. Rahman, Z. Begum, M. Z. Hafiz and A. A. Mahmud, "Synthesis of fault tolerant reversible logic circuits", In Proc. IEEE International Conference on Testing and Diagnosis, Chengdu, China, 28-29 April, 2009.
[10] M. S. Islam and Z. Begum, "Reversible logic synthesis of fault tolerant carry skip BCD adder", Bangladesh Academy of Science Journal, vol. 32, no. 2,pp. 193-200, 2008.
[11] R. Feynman, "Quantum mechanical computers", Optical News, vol. 11, 1985, pp. 11-20.
[12] E. Fredkin and T. Toffoli, "Conservative logic", Intl. Journal of Theoretical Physics, pp. 219-253, 1982.
[13] T. Toffoli, "Reversible computing", In Automata, Languages and Programming, Springer-Verlag, pp. 632-644, 1980.
[14] A. Peres, "Reversible logic and quantum computers", Physical Review: A, vol. 32, no. 6, pp. 3266-3276, 1985.
[15] J. W. Bruce, M. A. Thornton, L. Shivakumaraiah, P.S. Kokate, X. Li,  "Efficient adder circuits based on a conservative reversible logic gates", In Proceedings of IEEE Computer Society Annual Symposium on VLSI, Pittsburg, PA, pp. 83-88, 2002.
[16] M. Haghparast and K. Navi, "A novel fault tolerant reversible gate for nanotechnology based systems", Am. J. of App. Sci., vol. 5, no.5, pp. 519-523, 2008.
[17] M. S. Islam, M. M. Rahman, Z. Begum, M. Z. Hafiz, "Fault Tolerant reversible logic synthesis: carry look-ahead and carry-skip adders", In Proc. of IEEE International Conference on Advances in Computational Tools for Engineering Applications, pp. 396-401, July 15-17, 2009.

**Md. Saiful Islam** is Lecturer in the Institute of Information Technology at the University of Dhaka., Dhaka-1000, Bangladesh. He holds MS degree in Computer Science and Engineering from the University of Dhaka, 2007. Mr. Islam has more than 22 research articles published in several journals and conferences. Mr. Islam also leads and teaches modules at both BSc and MSc levels in Information Technology and Software Engineering. His research interests include Reversible Logic Design, Multimedia Information Retrieval, Machine Learning, Evolutionary Computing and Pattern Recognition.

**Muhammad Mahbubur Rahman** received his master degree in Information Technology from Institute of Information Technology at University of Dhaka. Currently, he is working as a lecturer in the department of Computer Science at American International University-Bangladesh. His research interests include Data Mining, Machine Learning, Bioinformatics, Game Theory, Artificial Intelligence, Econometrics and Logic Circuit Minimization. Currently his active research works are in bioinformatics and data mining. He has several years of working experience in local and international telecommunication and software companies. He is a founder and developer of a famous personalized search engine, www.vabuk.com.

**Dr. Zerina Begum** obtained her B. Sc (Hons) and M. Sc in Physics from the University of Dhaka in1986 and 1987 respectively. Dr. Begum started her carrier as a Research Fellow in Bose Research Centre at University of Dhaka. Two years later she joined as a Programmer in the Computer Centre of the same university. She was awarded M. Phil for her research in Semiconductor Technology and Computerized Data Acquisition System in 1995. She was repositioned as a Lecturer in the centre in 1997. As recognition of her outstanding research in organic semiconductor, Department of Physics, University of Dhaka awarded her with Ph. D. degree in 2007. She is still continuing to serve University of Dhaka as a Professor in the Institute of Information Technology. Her current research interest areas include Semiconductor Technology, Computerized Data Acquisition System, Advanced Digital Electronics, Software Engineering and Computational Biology. Dr. Begum is looking forward to build up IIT as the vanguard of ICT education and research in South Asia.

**Mohd. Zulfiquar Hafiz** obtained his B. Sc (Hons) and M. Sc in Pure Mathematics from the University of Dhaka in 1986 and 1987 respectively. Mr. Hafiz started his carrier as a programmer in the Computer Centre at University of Dhaka in 1993. He was repositioned as a Lecturer in the centre in 1997. He is still continuing to serve University of Dhaka as an Associate Professor in the Institute of Information Technology. His current research interest areas include Computer Networks, Embedded Systems and Computational Fluid Dynamics.